# First and High Order Sliding Mode-Multimodel Stabilizing Control Synthesis using Single and Several Sliding Surfaces for Nonlinear Systems: Simulation on an Autonomous Underwater Vehicles (AUV)

*Ahmed RHIF, Zohra KARDOUS, Naceur BEN HADJ BRAIEK*
L.S.A
Laboratoire des Systèmes Avancés
Ecole Polytechnique de Tunisie
BP 743, 2078 La Marsa, Tunisia
Email: ahmed.rhif@issatso.rnu.tn, zohra.kardous@enit.rnu.tn, naceur.benhadj@ept.rnu.tn

*Abstract* ─ This paper provides new analytic tools for a rigorous control formulation and stability analysis of sliding mode-multimodel controller (SM-MMC). In this way to minimise the chattering effect we will adopt as a starting point the multimodel approach to change the commutation of the sliding mode control (SMC) into fusion using a first order then a high order sliding mode control with single sliding surface and, then, with several sliding surfaces. For that the stability conditions invoke the existence of two Lyapunov-type functions, the first associated to the passage to the sliding set in finite time, and the second with convergence to the desired state. The approaches presented in this work are simulated on the immersion control of a submarine mobile which presents a problem for the actuators because of the high level of system non linearity and because of the external disturbances. Simulation results show that this control strategy can attain excellent performances with no chattering problem and low control level.

*Keywords*: Sliding mode, Multimodel, fusion, chattering, Lyapunov, stabilization.

## 1. Introduction

Variable structure systems (VSS) and sliding mode control (SMC) theory have proven effectiveness through the reported theoretical studies thanks to its robustness with respect to parameter variations and external perturbations. Its principal scopes of application are robotics and the electrical engines [1-5]. Sliding mode control systems which are a particular case of the variable structure systems are closed loop systems with discontinuous control that switch the system structure in order to maintain its trajectory inside the sliding surface. However, these performances are obtained at the price of some disadvantages. Indeed, to ensure the convergence of the system to the wished state, a high level switching control is often requested, this one may generate the chattering phenomenon which can be harmful for the systems' actuators. In this field, the multimodel approach constitutes a powerful tool for the identification, the control and the analysis of the complex systems. The principle of the multimodel representation makes possible to design a non linear control composed by the linear controls associated with each model. The global control can be then deduced either by a fusion or by a commutation between the different partial controls. The control by sliding mode multimodel (SM-MM) is inspired from the controls designed in [6-10]. In this way, a non linear system represented by linear sub models and associated to a sliding surface is considered in [6]. The process is chosen by a commutation between these different sub models weighted by adapted validities.

In addition in [7], the author considers the design of a non linear control system for an unmanned combat air vehicle for executing agile manoeuvres over the full flight envelope. The smooth aerobatic and complex combat manoeuvres are decomposed into a specific set of different sub manoeuvres to cover any arbitrary flight movement. To control each sub mode an inner/outer control loop approach with higher order sliding mode controllers are developed. To avoid the chattering phenomenon and the disturbances, fuzzy mode was applied. These controllers attain robust tracking of manoeuvre profiles for non linear aircraft dynamics. Resulting algorithms are applied to a high fidelity six degrees of freedom F-16 fighter aircraft model.

In another hand, an important problem in the field of the nonlinear systems is the search for stability criteria. For that to improve the quality of the control, we have to guarantee not only the stability of the system but also the means of stabilization [8-10]. In this way, before determine the fields of stability, some fundamental concepts of the stability theory will be recalled. In fact, many theories establish the fact that the systems which trajectories are attracted towards a balance point are asymptotically stable and lose energy gradually in a monotonous way. Hence, Lyapunov generalizes concept of energy while using quadratic or candidate functions *V(X)* which depends on the system state. For stabilization, we have to accomplice two important tasks for the commutated linear systems: the search for a commutation law of stabilizing and the synthesis of correctors stabilizing the system independently of the commutation law [11].

This paper is organized in four parts: first we begin by modelling the robotic process: a submarine mobile. Second, we present the sliding mode approach. After that, we introduce the SM-MM control which combines two approaches: sliding mode and multimodel approaches. Finally, we expose the experimental results.

## 2. Process modelling

The Autonomous Underwater Vehicles (AUVs) can be indexed in two classes depending on the immersion depth. We will speak then about AUVs coastal and AUVs deep seas. From a few hundred meters of depth, the dimensions structure and the AUVs characteristics change. This limit of depth will separate the vehicles deep seas from the coastal vehicles.

Today, the underwater robots are an integral part of the scientific equipment for seas and ocean exploration. Many examples showed that ROVs (Remotely Operating Vehicles) and AUVs (Autonomous Underwater Vehicles) are used in many fields and this for various applications like the inspection, the cartography or bathymetry.

However, we can distinguish a limiting depth for the various types of existing autonomous underwater machines. Indeed, starting from 300 meters, the structure, dimensions and the characteristics of these vehicles change. We have, on a side, AUVs Hugin 3000 type of Kongsberg Simrad, the Sea Oracle of Bluefin Robotics or Alistar 3000 of ECA, which can reach depths of 3000 meters, have a very great autonomy, considerable dimensions and a weight which requires an important logistics. On another side, AUVs of Remus Hydroid or Gavia Hyfmind types, with much less autonomy, but of reduced dimensions and logistics and with good modularity capacities that seems to be the perfect tool for the exploration of not very deep water.

In this context, the LIRMM and the Eca-Hytec company became partners to develop the first prototype of the AUV H160. This prototype was developed to surf and position with the using a GPS. On surface, the torpedo must be able to transmit the mission's data. The applications concerned are the inspection, bathymetry, the chemical data acquisition or sonar and video images. The machine will have also the possibility of surfing between 1 and 2 meters of depth with quasi no angle of pitching.

H160 is a torpedo type vehicle of a small size and of a low costs dedicated to the applications on not very deep water (up to 160 meters). The vehicle measures 1,80m length for a diameter of 20cm and a weight of 50kg. Thanks to its small size, the tests on the sea require a logistics reduced to the minimum to two people and a motor boat. The prototype is able to accomplish a mission of at least three hours with maintaining its speed with 3 knots. Its positive floatability makes possible that the torpedo goes back to surface after each end of mission. H160 is fed by a battery 48V/16Ah of the NiMH type, has an actuator with D.C current 230W and 430N.cm servo-motors for the riders control. The torpedo immersion capacity with no angle of pitching is due to its pair of surface riders that constitutes the main feature of this machine [12-14].

The torpedo is a cylindrical vehicle form as shown on Figure 2. Its structure is mainly made up of aluminium. We can detail the prototype in seven parts:

1. The principal part is the electronic section, composed of two stages. The first stage accommodates the battery, while the second one is composed of all the embarked charts (sensors, power, PC,…). This part is obviously tight;
2. Section made by the antennas GPS, Radio and Wifi, also by the riders control of the front immersion;
3. The sensor CTD and Sidescan sonar are in a wet part;
4. The Doppler Log is located in a tight part;
5. The nose of the vehicle composed by a camera CCD and two sounders;
6. Behind the principal part, we find the pressure pick-ups and an emergency acoustic pinger in a wet part;
7. Finally, the propeller and the riders constitute the engine back part.

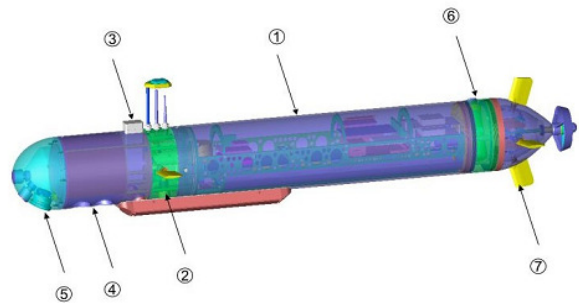

**Fig. 2 Different components of the H160**

Submarine mobiles present strong non linearity and always subject to disturbances and parameters uncertainties which make their measurement and their control a hard task and may present a harmful effect to actuators. The Autonomous Underwater Vehicle (AUV) named H160 (Fig.1.) represented in is a torpedo type robots of a small size; it is usually used in little deep water; until 160m. The vehicle measures 1,90m length for 15cm of diameter and weights 40kg. To model this system, we need to define two referentials [15-17]: one fix referential related to the vehicle which is defined in an origin point: $R_0$ ($X_0$, $Y_0$, $Z_0$) and the second related to the Earth $R(x, y, z)$.

The cinematic model is represented as follows:

$$\eta = J_c(\eta_2)v = \begin{bmatrix} \dot{\eta}_1 \\ \dot{\eta}_2 \end{bmatrix} = \begin{bmatrix} J_{c1}(\eta_2) & 0_{3x3} \\ 0_{3x3} & J_{c2}(\eta_2) \end{bmatrix} \begin{bmatrix} v_1 \\ v_2 \end{bmatrix} \quad (1)$$

with $J_{c1}(\eta_2)$ is the linear speed transformation matrix and $J_{c2}(\eta_2)$ is the angular speed transformation matrix.

$$\eta = \begin{bmatrix} \eta_1 \\ \eta_2 \end{bmatrix}; \quad \eta_1 = \begin{bmatrix} x \\ y \\ z \end{bmatrix}; \quad \eta_2 = \begin{bmatrix} \phi \\ \theta \\ \psi \end{bmatrix} \quad (2)$$

$$v = \begin{bmatrix} v_1 \\ v_2 \end{bmatrix}; \quad v_1 = \begin{bmatrix} u \\ v \\ \omega \end{bmatrix}; \quad v_2 = \begin{bmatrix} p \\ q \\ r \end{bmatrix} \quad (3)$$

$$\Gamma = \begin{bmatrix} \Gamma_1 \\ \Gamma_2 \end{bmatrix}; \quad \Gamma_1 = \begin{bmatrix} X \\ Y \\ Z \end{bmatrix}; \quad \Gamma_2 = \begin{bmatrix} K \\ M \\ N \end{bmatrix} \quad (4)$$

η the state vector representing the robot position related to the R(x, y, z) reference.
ν represent the robot speed related to $R_0(x_0, y_0, z_0)$.
Γ the forces vector applied to the mobile.
ω is linear velocity, **q** the angular velocity, **θ** the angle of inclination and **z** the depth.
The dynamic equation is represented by:

$$M_\eta(\eta)\ddot{\eta} + C_\eta(\upsilon,\eta)\dot{\eta} + D_\eta(\upsilon,\eta)\eta + g_\eta(\eta) = \tau_\eta \quad (5)$$

with $M_\eta$ the inertia matrix, $C_\eta$ the Coriolis matrix, $D_\eta$ the matrix of rubbing forces, $g_\eta$ the hydrostatic effort vector and $\tau_\eta$ the input control vector.

In order to control the behavior of an underwater vehicle in the immersion phase, we must be able to vary its buoyancy. The buoyancy of a vehicle in immersion is the difference between the Archimede pressure and the gravity. Buoyancy (noted Φ1) depends on the vehicle mass (m), its volume (V) and the density of water (ρ). So we define (6):

$$\Phi 1 = \rho V - m \quad (6)$$

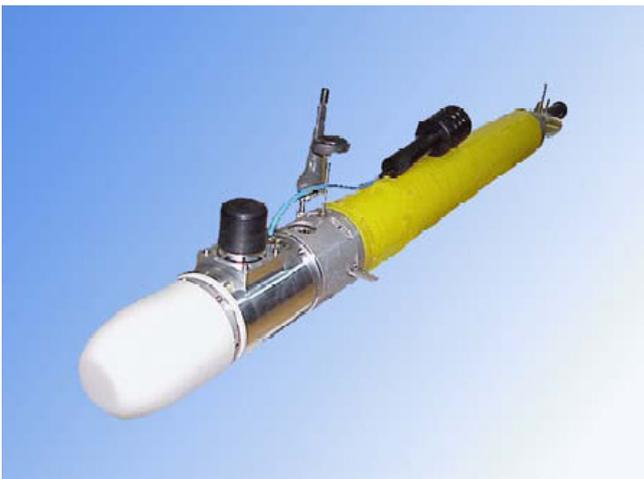

Fig.1. Submarine engine

In this study we will consider only the immersion variable. For that the depth *z* variation will be considered as follow:

$$0.7 \leq z \leq 1.3$$

The AUV present a strong nonlinear aspect that appears when we describe the system in 3 dimensions (3D), so the state function will present a new term of disturbances. The state space describing the system is given below:

$$\dot{x} = Ax + Bu + \varphi(x,u) \quad (7)$$

with,

$$\dot{X} = \begin{bmatrix} \dot{\omega} \\ \dot{q} \\ \dot{\theta} \\ \dot{z} \end{bmatrix}, \quad A = \begin{bmatrix} 0.47 & 0.3 & 0 & 0 \\ -0.69 & 0.79 & 0.36 & 0 \\ 0 & 1 & 0 & 0 \\ 1 & 0 & 1 & 0 \end{bmatrix} \text{ and } B = \begin{bmatrix} 0.05 \\ 0.14 \\ 0 \\ 0 \end{bmatrix}$$

$$\|\varphi(x,u)\| < Mx$$

To control such system we start by applying the sliding mode control which is characterized by its robustness and effectiveness.

### 3. The sliding mode control

The sliding mode control consists in bringing back the state trajectory towards the sliding surface and to make it move above this surface until reaching the equilibrium point. If, for initial state vector $x(t_0) \in S$, the state trajectory remains in the hypersurface $S_i$, $x(t) \in S \; \forall t > t_0$, then $x(t)$ obeys to the sliding mode of the system.

*3.1 A sliding mode control synthesis*
There are three different sliding mode structures: in the first one, commutations take place on the control unit, the second structure uses commutations on the feedback state and in the last one, the commutations occur on the control unit with addition of the equivalent control. In this study we adopt the last structure because it is the most solicited (Fig.2.).

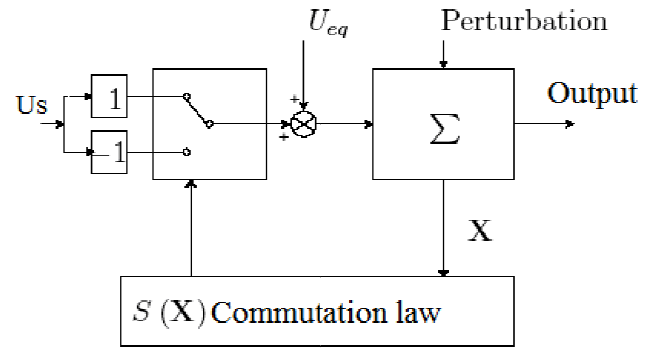

Fig.2. Control unit with addition of the equivalent control

To ensure the existence of the sliding mode, first, we must produce a high level commutation control. This property can be applied by a relay which commutates between two extreme values $u_s = \pm A$ and that gives the desired result when *A* is sufficiently high. This switching control can be represented by many forms and may satisfy the stability condition: $s\dot{s} < 0$

Here we choose the control forms (10). Second, we have to define a sliding surface. In this case, we consider a linear sliding surface (9).

$$s = CX = 0 \quad (9)$$
$$u_s = -k|s|sign(s) = -ks \quad (10)$$

with k>0.

To compute the gain *k* that makes the system stable in the convergence phase to the sliding surface, we choose a quadratic Lyaponov function $V = \frac{1}{2}s^2$ and we have to prove that $\dot{V} = s\dot{s} < 0$.

We start by calculating the first derivative of the considered sliding surface (9) as below:

$$\dot{s} = CAx + CBu + C\varphi(x,u) \quad (11)$$

as we have:

$$-Mx < \varphi(x,u) < Mx \quad then \quad -CMx < C\varphi(x,u) < CMx$$

$$so \quad s\dot{s} < s(CAx + CBu + CMx)$$

Now, let's search u satisfying $s(CAx + CBu + CMx) < 0$

Knowing that in the convergence phase we have $u_s \approx u$

Equations (9), (11) and (10) give:

$$\Leftrightarrow x^T \left( C^T CA - C^T CBkC + C^T CM \right) x < 0$$

$$\Leftrightarrow k > (C^T CBC)^{-1} \left( C^T CA + C^T CM \right)$$

Then, to make the system converge to the sliding surface, we have to ensure the equation (12) that guarantees $s\dot{s} < 0$.

$$k > (BC)^{-1}(A + MI) \quad (12)$$

with *I* the identity matrix

In the reaching phase, we note that $u_{eq} \approx u$.

To compute the stabilizing control law of the system (7), we use the fact that $s\dot{s} < 0$.

as we have: $CAx + CBu - CMx < \dot{s} < CAx + CBu + CMx$

Then In the reaching phase to the desired state, the following system (7) gives the stabilizing control law of the system.

$$u = \begin{cases} -(CB)^{-1}(CA - CM)x - \varepsilon & if \quad s > 0 \\ -(CB)^{-1}(CA + CM)x + \varepsilon & if \quad s < 0 \end{cases} \quad (13)$$

with $\varepsilon > 0$ and *I* the identity matrix.

The simulation result (Fig.3), of this first order sliding mode control, on the submarine mobile shows that we can reach the desired value of depth (Fig.3-a) in a short time (10s), Other ways, we notice that the steady state present some oscillations. However, the control level (*u=2*) and the switching frequency are high (Fig.3-b).

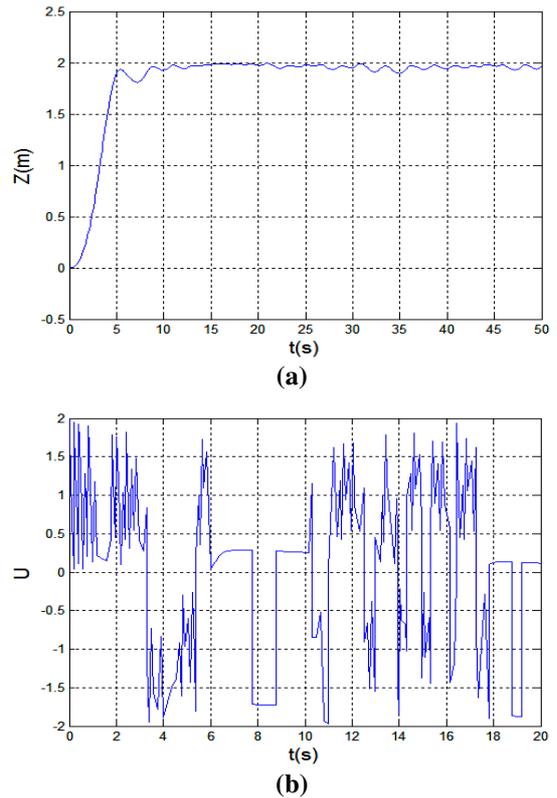

Fig.3. System evolution by first order sliding mode control

For any control device which presents non linearity such as delay or hysteresis, limited frequency commutation is often imposed, other ways, the state oscillation will be preserved even in vicinity of the sliding surface. This behaviour is known by chattering phenomenon. This phenomenon can cause damage to actuators or to the plant itself. As a solution, the high order sliding mode control is much solicited.

*3.2 High order sliding mode control*
The high order sliding mode control consists in considering the derivatives of the sliding variable. This method allows the rejection of the chattering phenomenon while preserving the robustness of the approach [18-19].

$$S^r = \left\{ x \in IR^n : s = \dot{s} = ... = s^{(r-1)} = 0 \right\}, r \in IN$$

$r \geq \rho$, $\rho > 0$, *s(x,t)* the sliding function : which is a differentiable function with its *(r - 1)* first time derivatives depending only on the state *x(t)*.

In the case of second order sliding mode control, the following relation must be verified:

$$s(t,x) = \dot{s}(t,x) = 0 \quad (14)$$

The derivative of the sliding function is

$$\frac{d}{dt}s(t,x) = \frac{\partial}{\partial t}s(t,x) + \frac{\partial}{\partial x}s(t,x)\frac{\partial x}{\partial t} \quad (15)$$

Considering relation (14) the following equation can be written:

$$\dot{s}(t,x,u) = \frac{\partial}{\partial t} s(t,x) + \frac{\partial}{\partial x} s(t,x) \dot{x}(t,u) \quad (16)$$

The second order derivative of $S(t,x)$ is :

$$\frac{d^2}{dt} s(t,x,u) = \frac{\partial}{\partial t} \dot{s}(t,x,u) + \frac{\partial}{\partial x} \dot{s}(t,x,u) \frac{\partial x}{\partial t} + \frac{\partial}{\partial u} \dot{s}(t,x,u) \frac{\partial u}{\partial t} \quad (17)$$

This last equation can be written as follows:

$$\frac{d}{dt} \dot{s}(t,x,u) = \xi(t,x) + \psi(t,x) \dot{u}(t) \quad (18)$$

with:

$$\xi(t,x) = \frac{\partial}{\partial t} \dot{s}(t,x,u) + \frac{\partial}{\partial x} \dot{s}(t,x,u) \dot{x}(t) \quad (19)$$

$$\psi(t,x) = \frac{\partial}{\partial u} \dot{s}(t,x,u) \quad (20)$$

We consider a new system whose state variables are the sliding function $s(t,x)$ and its derivative $\dot{s}(t,x)$.

$$\begin{cases} \omega_1(t,x) = s(t,x) \\ \omega_2(t,x) = \dot{s}(t,x) \end{cases} \quad (21)$$

Equations (18) and (21) lead to:

$$\begin{cases} \dot{\omega}_1(t,x) = \omega_2(t,x) \\ \dot{\omega}_2(t,x) = \xi(t,x) + \psi(t,x) \dot{u}(t) \end{cases} \quad (22)$$

Equations (20) and (23) lead to:

$$\begin{cases} \dot{\omega}_1(t,x) = \omega_2(t,x) \\ \dot{\omega}_2(t,x) = \xi(t,x) + \psi(t,x) \dot{u}(t) \end{cases} \quad (23)$$

In this way a new sliding function $\sigma(t,x)$ is proposed:

$$\sigma(t,x) = \omega_2(t,x) + \alpha\, \omega_1(t,x) = \dot{s}(t,x) + \alpha s(t,x) \quad (24)$$

with $\alpha > 0$.

When applying the second order sliding mode control to the system (7), we obtain results presented in figure 4. The Fig.4-a shows that the high order sliding mode control can reduce considerably the chattering phenomenon but the level and the commutation frequency of the control are always high (Fig.4-b).

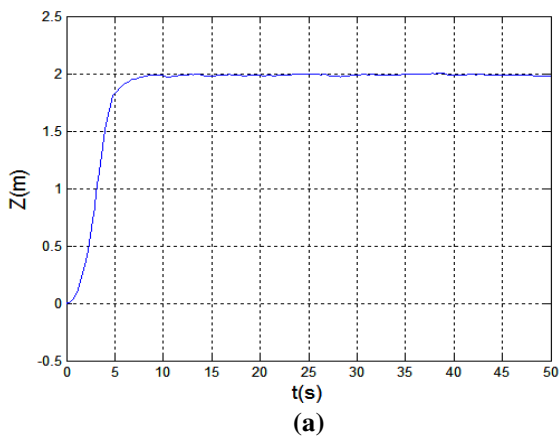

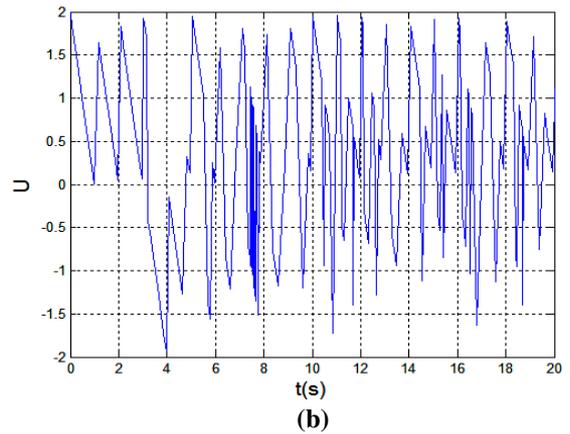

Fig.4. System evolution by second order sliding mode control

### 4. The sliding mode multimodel control

In fact, our approach consists in carrying out a fusion on the sliding mode control instead of a commutation in order to eliminate or minimize the oscillations on the sliding surface. The equivalent control resulted by this operation will control the process as shown in Fig.5.

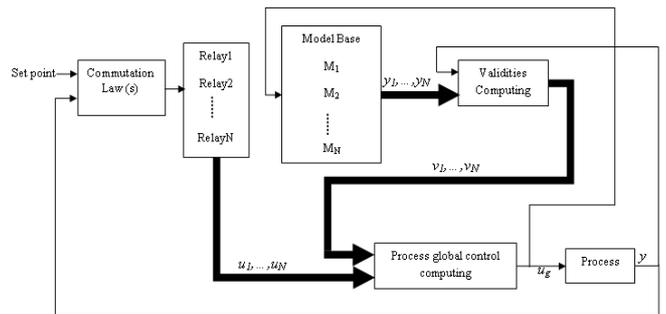

Fig.5. Sliding mode multimodel control structure

*4.1 The Multimodel approach*

The multimodel approach represents an interesting alternative and a powerful tool in the identification, the control and the analysis of complex systems. Consider system (25).

$$\begin{cases} \dot{x} = A_i x + B_i u \\ y = C_i x \end{cases} \quad (25)$$

The multimodel control in which we will be interested consists in the fusion of partial controls [20]. For that we have to compute the validity of each partial model and associate the sub controls weighted by the correspondent coefficients. The obtained result will control the global process (26).

$$u(t) = \sum_{i=1}^{N} v_i(t) u_i(t) \quad (26)$$

Then the system will be presented as follows (27).

$$\begin{cases} \dot{x} = \sum_{i=1}^{n} v_i (A_i x + B_i u) \\ y = \sum_{i=1}^{n} v_i C_i x \end{cases} \quad (27)$$

with $v_i$, $i=1,...,n$, the correspondent validities. These weighting coefficients must satisfy the convex sum property (28).

$$\begin{cases} v_i \in [0,1], \quad i = 1....N \\ \sum_{i=1}^{N} v_i = 1 \end{cases} \quad (28)$$

Several methods of validities calculation were already presented in the literature [21-24]. The common one is the residue approach.

$$v_i(t) = 1 - r_{in}(t) \quad (29)$$

where

$$r_{in}(t) = \frac{r_i(t)}{\sum_{i=1}^{N} r_i(t)} \quad (30)$$

and

$$r_i(t) = |y(t) - y_i(t)|; \; i=1,...,N \quad (31)$$

$v_i$ is the validity, $r_i$ the residue, $y(t)$ is the system's output, $y_i(t)$ is the output of the $i^{th}$ model.

In order to reduce the perturbation phenomenon due to the inadequate models, we reinforce the validities as follows (32):

$$v_i^{renf}(t) = v_i(t) \prod_{\substack{j=1 \\ i \ne j}}^{N} (1 - v_j(t)) \quad (32)$$

The normalized reinforced validities are given by (33).

$$v_{in}^{renf}(t) = \frac{v_i^{renf}(t)}{\sum_{i=1}^{N} v_i^{renf}(t)} \quad (33)$$

Consider the free regime system; the system will be represented by (34).

$$\dot{x} = \sum_{i=1}^{n} v_i A_i x \quad (34)$$

The system is stable if there exists $P > 0$ a symmetric matrix that makes the first derivative of Lyaponov quadratic equation negative: $\dot{V}(x) < 0$ [25].

For this case, we chose the quadratic Lyapunov function: $V(x) = x^T P x$

$$\dot{V}(x) = \dot{x}^T P x + x^T P \dot{x} = (\sum_{i=1}^{N} v_i x^T A_i^T) P x + x^T P (\sum_{i=1}^{N} v_i A_i x)$$

$$= \sum_{i=1}^{N} v_i (x^T A_i^T P x + x^T P A_i x) = x^T (\sum_{i=1}^{N} v_i (A_i^T P + P A_i)) x$$

Then, the stability condition adopting the fusion approach is respected when $\dot{V}(x) < 0$ which gives (35).

$$\begin{cases} P > 0 \\ A_i^T P + P A_i < 0 \end{cases} \quad (35)$$

with $P > 0$

In the case of a state feedback control the system (27) will be as represented follow:

$$\dot{x} = \sum_{i=1}^{n} \sum_{j=1}^{n} v_i v_j (A_i - B_i k_j) x \quad (36)$$

The partial control is $u_i = -k_i x$ with $k_i > 0$ and the global control of the process is $u = \sum_{i=1}^{n} v_i u_i$.

we take : $G_{ij} = A_i - B_i k_j$

this will make :

$$\dot{x} = \sum_{i=1}^{n} \sum_{j=1}^{n} v_i^2 G_{ii} x + 2 \sum_{i<j} v_i v_j \left( \frac{G_{ij} + G_{ji}}{2} \right) x$$

To verify the stability condition, we evaluate the first derivative of the Lyaponov function $V(x) = x^T P x$

$$\dot{V}(x) = \dot{x}^T P x + x^T P \dot{x}$$

$$= \left( \sum_{i=1}^{n} v_i^2 G_{ii}^T x^T + 2 \sum_{i<j} v_i v_j \left( \frac{G_{ij}+G_{ji}}{2} \right)^T x^T \right) P x + x^T P \left( \sum_{i=1}^{n} v_i^2 G_{ii} x + 2 \sum_{i<j} v_i v_j \left( \frac{G_{ij}+G_{ji}}{2} \right) x \right)$$

$$= x^T \left\{ \sum_{i=1}^{n} v_i^2 (G_{ii}^T P + P G_{ii}) + 2 \sum_{i<j} v_i v_j \left[ \left( \frac{G_{ij}+G_{ji}}{2} \right)^T P + P \left( \frac{G_{ij}+G_{ji}}{2} \right) \right] \right\} x$$

The stability condition of the system (27) adopting the fusion approach is verified when $\dot{V}(X) < 0$ that gives following conditions (37).

$$\begin{cases} G_{ii}^T P + P G_{ii} < 0 \\ \left( \frac{G_{ij}+G_{ji}}{2} \right)^T P + P \left( \frac{G_{ij}+G_{ji}}{2} \right) < 0 \end{cases} \quad (37)$$

*4.2 Formulation of the sliding mode multimodel approach*

In fact, the approach in which we interest in this paper consists in carrying out a fusion on the sliding mode control instead of a commutation, as shown in Fig.1, in order to eliminate or minimize the chattering phenomenon. To synthesize the global SM-MMC of the process we have to respect the following algorithm:

First, we start by fixing the different models $M_i$ $i=1,..,n$ relative to the different balance points or the extreme models. Then we have to choose switching controls $u_{si}$ ($i=1,..,n$) of relays type (38). The partial controls $u_i$ ($i=1,..,n$) (39) are obtained by adding the equivalent control $u_e$ to $u_{si}$. The global control $u_g$ of the process will be deduced by summing the partials controls $u_i$ weighted by the correspondent validities $v_i$ computed on line (40).

$$u_{si} = \begin{cases} u_{si\min} & if \quad sign(s) < 0 \\ u_{si\max} & if \quad sign(s) > 0 \end{cases} \quad (38)$$

with $u_{si} = -k_i|s|sign(s) = -k_i s$, i=1,…,N.

$$u_i = u_e + u_{si} \quad (39)$$

$$u_g = \sum_{i=1}^{N} v_i u_i \quad (40)$$

$u_e$ the equivalent control.

In this section we will try to synthesise a SM-MMC for the submarine mobile and to identify the stabilizing conditions of this control. First, we start by the case of single sliding surface (42).
Knowing that in the sliding surface the systems' order is reduced, we consider:

$$\begin{aligned} s &= x_n + \alpha_{n-1}x_{n-1} + \cdots + \alpha_1 x_1 = 0 \\ x_n &= -\alpha_{n-1}x_{n-1} - \cdots - \alpha_1 x_1 = -L_{n-1}X_{n-1} \end{aligned} \quad (41)$$

with $L_{n-1} = [\alpha_1 \ \alpha_2 \cdots \alpha_{n-1}], \alpha_i > 0$, $X_{n-1} = \begin{pmatrix} x_1 \\ \vdots \\ x_{n-1} \end{pmatrix}$

We take :

$$s_i : x_n = -L_{n-1}^i X_{n-1} \quad (42)$$

### 4.2 Sufficient stabilizing conditions of SM-MMC

The sliding surface used in this study is given by (42).

**Theorem 1:** The asymptotic stability condition of the system (27) governed by SM-MMC using the sliding surface (42) is provided by (43) and (44).

$$k > (BC)^{-1}(A + MI) \quad (43)$$

$$\begin{cases} P > 0 \\ (A_{n-1}^i - B_{n-1}^i L_{n-1})^T P_{n-1} + P_{n-1}(A_{n-1}^i - B_{n-1}^i L_{n-1}^i) < 0 \end{cases} \quad (44)$$

where :

$$P = \begin{pmatrix} P_{n-1} & 0 \\ 0 & P_n \end{pmatrix}$$

*Proof-Theorem 1:* Consider the following Lyaponov quadratic function:

$$\begin{aligned} V(x_{n-1}) &= x_{n-1}^T P_{n-1} x_{n-1} \\ \dot{V}(x_{n-1}) &= \dot{x}_{n-1}^T P_{n-1} x_{n-1} + x_{n-1}^T P_{n-1} \dot{x}_{n-1} \\ &= x_{n-1}^T (A_{n-1}^i - B_{n-1}^i L_{n-1}^i)^T P_{n-1} x_{n-1} + x_{n-1}^T P_{n-1}(A_{n-1}^i - B_{n-1}^i L_{n-1}^i) x_{n-1} \\ &= x_{n-1}^T \left[ (A_{n-1}^i - B_{n-1}^i L_{n-1}^i)^T P_{n-1} + P_{n-1}(A_{n-1}^i - B_{n-1}^i L_{n-1}^i) \right] x_{n-1} \end{aligned}$$

To ensure that $\dot{V}(x_{n-1}) < 0$ we must verify the condition (44).

After the simulation of this control on the AUV mobile, the output evolution (Fig.6) shows a reduction of the chattering phenomenon in the case of the first order SM-MM relatively to that of the first order sliding mode. This amelioration is also noticed on the control level and the switching frequency. In terms of improvement of the system output and to have a more rapid convergence on the sliding function, we think to use a high order SM-MM control. In this case, the sliding surface used is defined in (42). The simulation results of the submarine system (7) are illustrated in figure 7.

The simulation shows that the high order SM-MM control is the best alternative to reduce considerably the chattering effect relatively to the three last approaches simulated in this paper. Moreover, we notice that the control level is lightly smaller than the SMC level and the commutation frequencies are always sharp which can be harmful to the submarine actuators. In this way, we think to use a SM-MMC with several sliding surfaces.

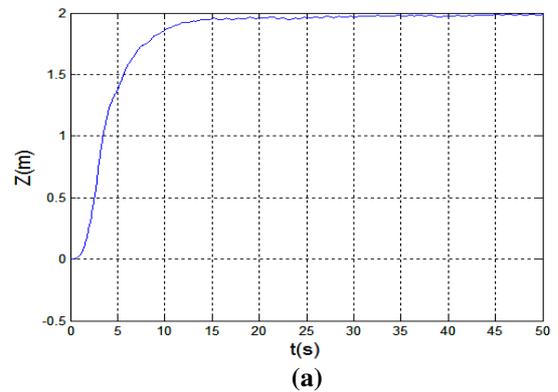

(a)

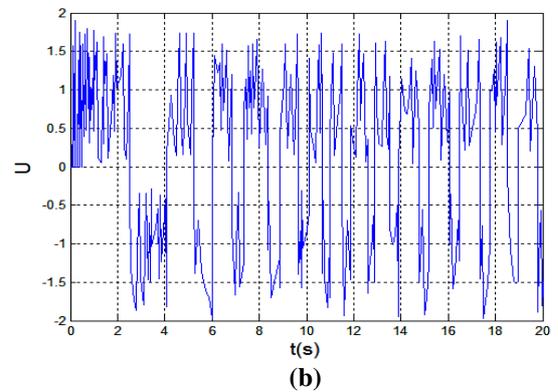

(b)

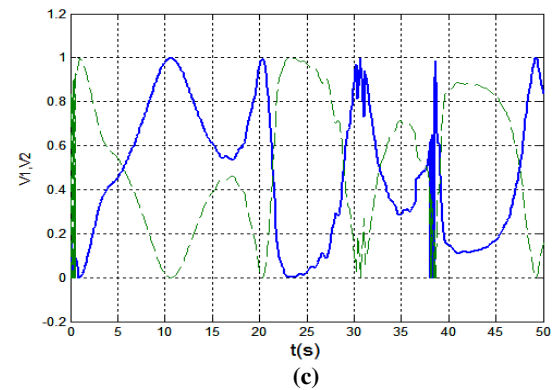

(c)

Fig.6. System evolution by first order SM-MM control

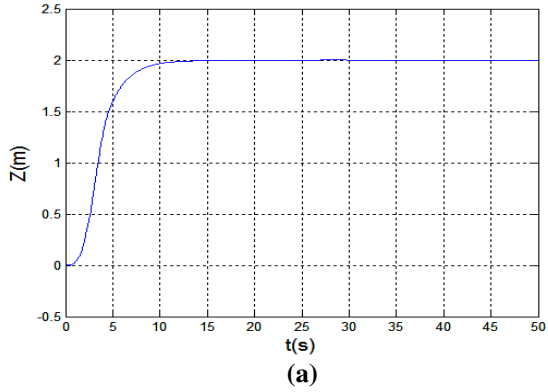
(a)

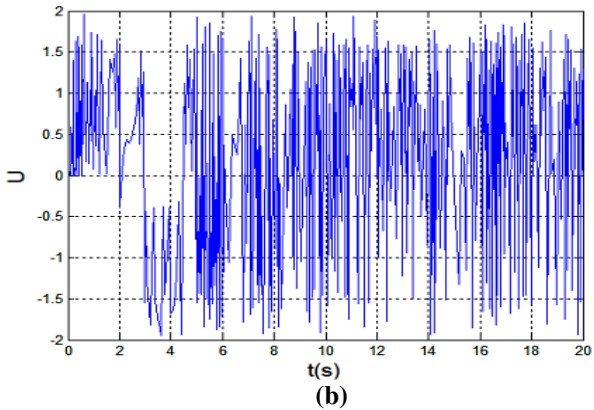
(b)

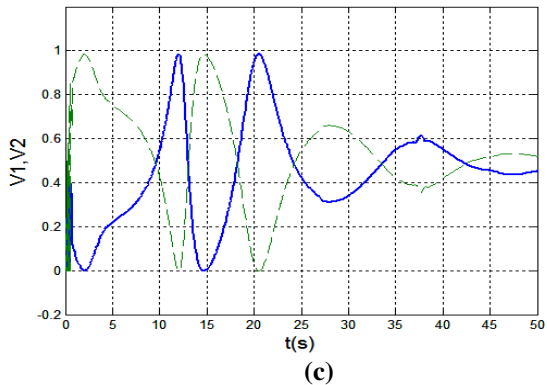
(c)

Fig.7. System evolution by second order SM-MM control

**5 Multi surfaces (SM-MM) approach performances**

Previous results have shown that multiplying the sliding mode order improve the system response quality. However, it causes high level commutation frequency in the control signal. In order to minimize the control discontinuities, we think about using multiple sliding surfaces while reducing the sliding mode order.
In this way, to improve the controlling process of each sub model, we think about using several first order sliding surfaces, each state of a sub model $M_i$ is considered to reach one of these sliding surfaces $s_i$ (Fig.8). To ensure the SMC existence, we use several switching control $u_{si}$ relative to each sliding surface $s_i$. Then, the partial control $u_i$ of each sub model will be computed as shown in (45). After that the process will converge to the sum of those surfaces weighted by the correspondent validities $v_i$ (46).

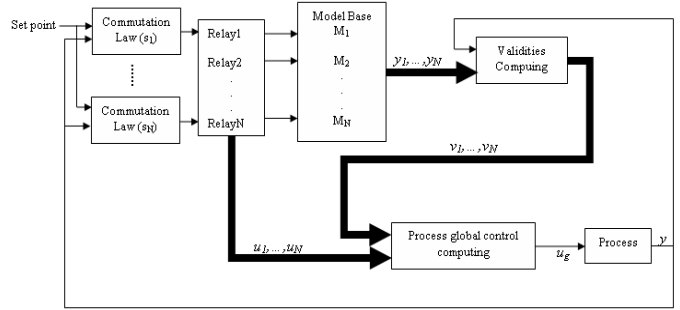

Fig.8. Sliding mode multimodel control structure (multi sliding surfaces)

$$S = \sum_i v_i s_i \quad (45)$$

$$u_i = u_{ei} + u_{si} \quad (46)$$

To verify the stability condition, we choose a non quadratic function operating in $s$ (47) and we have to verify that $\dot{V}(S) < 0$.

$$V(S) = \sum_{i=1}^{m} P_i s_i^2(x) \quad (47)$$

Noted that we use the fusion approach, the global process will be represented by (48).

$$\begin{cases} \dot{x} = \sum_{i=1}^{n} v_i(A_i x + B_i u + \varphi(x,u)) \\ y = \sum_{i=1}^{n} v_i C_i x \end{cases} \quad (48)$$

**Theorem 2**: The SM-MMC (3) stabilizes the system (20) if it fulfils the two conditions:
i)

$$\begin{cases} K > \sum_i \mu_i \left( (B_i T_i)^{-1} (A_i + MI) \right) \\ u = -\sum_i \left( v_i (B_i T)^{-1} (A_i + MI) x \right) - \varepsilon \end{cases} \quad (49)$$

with $0 < \mu_i, v_i < 1$, $T_i$ a linear vector and $\varepsilon > 0$.

ii)

$$\begin{cases} P_{n-1} > 0 \\ (A_{n-1}^i - B_{n-1}^i L_{n-1}^i)^T P_{n-1} + P_{n-1}(A_{n-1}^i - B_{n-1}^i L_{n-1}^i) < 0 \end{cases} \quad (50)$$

*Proof-Theorem 2:*
i) In the convergence phase we have to verify the condition $S\dot{S} < 0$ using a switching control $u_S = -KS$ with $K = \sum_i \mu_i k_i$

Equation (47) gives $\dot{V}(s) = \sum_{i=1}^m 2P_i s_i \dot{s}_i$

Consider that $s_i(x) = T_i x = 0$

$\dot{s}_i(x) = T_i \dot{x}$ we will have:

$\dot{s}_i(x) = T_i A_i x + T_i B_i u_i + T_i \varphi(x,u)$

$s_i \dot{s}_i = T_i x T_i A_i x + T_i x T_i B_i u + T_i x T_i \varphi(x,u)$

$= x^T T_i^T T_i A_i x + x^T T_i^T T_i B_i u + x^T T_i^T T_i \varphi(x,u)$

we use the fact that:

$\varphi(x,u) < Mx$ and $u_{si} = -k_i |s_i| sign(s_i) = -k_i s_i$

$\Rightarrow s_i \dot{s}_i < x^T \left[ T_i^T T_i A_i - T_i^T T_i B_i k_i T_i + T_i^T T_i MI \right] x$

$s_i \dot{s}_i < 0 \Rightarrow A_i - B_i k_i T_i + MI < 0 \Rightarrow k_i > (B_i T_i)^{-1}(A_i + MI)$

after fusion:

$$K > \sum_i \mu_i \left( (B_i T_i)^{-1}(A_i + MI) \right)$$

The explicit form of the control that make the system reach the sliding surface $S$ is given by the following equation (51).

$$u_i = -\mu_i (B_i T_i)^{-1}(A_i + MI)x - \varepsilon \quad (51)$$

Consider the non quadratic function operating in $s$ (47):

$V(s) = \sum_{i=1}^m P_i s_i^2(x) \Rightarrow \dot{V}(s) = \sum_i 2P_i s_i \dot{s}_i$

$\dot{V}(s) = \sum_i 2P_i \left( x^T T_i^T T_i A_i x + X^T T_i^T T_i B_i u + x^T T_i^T T_i \varphi(x,u) \right)$

$\dot{V}(s) < \sum_i 2P_i x^T T_i^T T_i (A_i x + B_i \mu_i u_i + Mx)$

then, $\dot{V}(s) < 0 \Rightarrow A_i x + B_i \mu_i u_i + Mx < 0 \Leftrightarrow u_i < -(\mu_i B_i)^{-1}(A_i + MI)x$

$\Leftrightarrow u_i = -(\mu_i B_i)^{-1}(A_i + MI)x - \varepsilon$

In this way, using (6), the global control is written as follow:

$u = -\sum_i \left( (v_i B_i)^{-1}(A_i + MI)x \right) - \varepsilon$

ii) In the reaching phase, we choose a non quadratic Lyaponov function:

$V(x_{n-1}) = \sum_i x_{n-1}^T P_{n-1} x_{n-1}$

$\dot{V}(x_{n-1}) = \sum_i \left( \dot{x}_{n-1}^T P_{n-1} x_{n-1} + x_{n-1}^T P_{n-1} \dot{x}_{n-1} \right)$

So $\dot{x}_{n-1}^T P_{n-1} x_{n-1} + x_{n-1}^T P_{n-1} \dot{x}_{n-1} < 0$ when $\dot{V}(x_{n-1}) < 0$.

Using Proof-Theorem 1 we got the condition (50).

The simulation results (fig.9) of this approach (multi surfaces) show that the system reach the desired state in a short time [0, 20s] with no chattering phenomenon. The control level and the switching frequency are less than the other approaches.

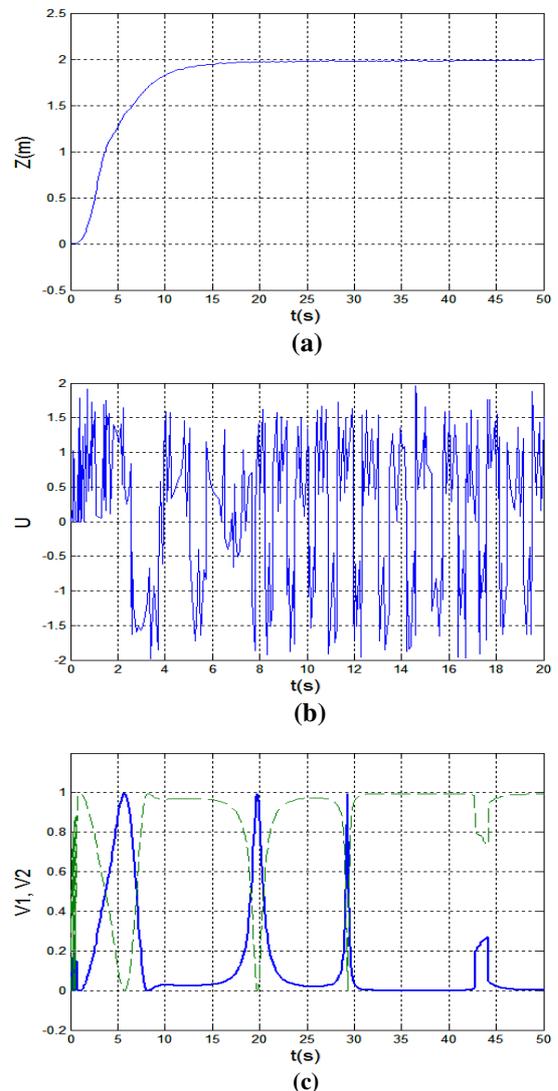

(a)

(b)

(c)

Fig.9. System evolution by first order SM-MM multi surfaces

## Conclusion

The simulation results of the considered system (submarine) illustrate the contribution of the sliding mode multimodel control (SM-MMC) as a stabilizing control law for nonlinear systems. The control law adopted was, first, of state feedback type, then, using the global SM-MMC. This study is based on Lyaponov theory: a quadratic and non quadratic criterion was developed. First we start by studying a system using only one sliding surface for that we choose a quadratic Lyaponov function, in the case of multi surfaces we use a non quadratic one. The sufficient conditions of stabilization are developed for a closed loop system with separable nonlinearity. Indeed, the multimodel fusion minimizes the oscillations in the output of the actuators by reducing considerably the level of the control and the chattering phenomenon. We notice that the SM-MMC conserves the sliding mode proprieties of robustness and rapidity.

We conclude that the first order SM-MMC with multi surfaces gives the best simulation results relatively to the other approaches presented in this paper. Notice that the derivative function is very hard to implant experimentally, this last approach could be also an easier approach for control application.